\documentclass[journal]{IEEEtran}
\usepackage{courier}
\usepackage{color,url,xspace}
\usepackage{amssymb,amsmath}
\usepackage{graphicx}
\usepackage{colordvi}
\usepackage{epsfig}
\usepackage{endnotes}
\usepackage{verbatim}
\usepackage{subfigure}
\usepackage{textcomp}
\usepackage {subfig}
\usepackage{setspace}
\begin{document}
\title{Joint Scalable Coding and Routing for 60\,GHz Real-Time Live HD Video Streaming Applications}

\author{Joongheon~Kim,~\IEEEmembership{Member,~IEEE, }
	Yafei~Tian,~\IEEEmembership{Member,~IEEE, }
	Stefan~Mangold,~\IEEEmembership{Member,~IEEE, }
	Andreas~F.~Molisch,~\IEEEmembership{Fellow,~IEEE}
\thanks{Parts of this paper were presented at the 22nd IEEE International Symposium on Personal Indoor and Mobile Radio Communications (PIMRC), Toronto, Canada, 14 September 2011 (refer to~\cite{kim11pimrc}).}
\thanks{J. Kim is with the Department of Computer Science, University of Southern California, Los Angeles, CA 90089, USA e-mail: joongheon.kim@usc.edu.}
\thanks{Y. Tian is with the School of Electronics and Information Engineering, Beihang University, Beijing 100191, China e-mail: ytian@buaa.edu.cn.}
\thanks{S. Mangold is with the Disney Research Zurich and the Laboratory for Software Technology, ETH Zurich, 8092 Zurich, Switzerland
e-mail: stefan@disneyresearch.com.}
\thanks{A.F. Molisch is with the Ming Hsieh Department of Electrical Engineering, University of Southern California, Los Angeles, CA 90089, USA e-mail: molisch@usc.edu.}
}
\maketitle

\begin{abstract}
Will be submitted later.
\end{abstract}
\begin{keywords}
\,60\,GHz, Multi-Gbit/s HD Video Streaming, Wireless Video Quality Maximization, Routing
\end{keywords}

\section{Introduction}\label{sec:intro}
Will be submitted later.


\end{document}